\documentclass[10pt]{wlscirep}
\usepackage{blindtext}
\usepackage{multicol}

\usepackage{graphicx}
\usepackage[english]{babel}
\usepackage{amsmath}
\usepackage{amsfonts}
\usepackage{hyperref}
\usepackage{cleveref}
\usepackage{booktabs}
\usepackage{float}
\usepackage{color}
\usepackage{xcolor}

\usepackage{subcaption}
\usepackage{geometry}
\usepackage{tikz}
\usepackage{amsmath,mathtools,amssymb}
\usetikzlibrary{calc}
\usepackage{caption}
\usepackage[utf8]{inputenc}
\usepackage[T1]{fontenc}

\definecolor{revised}{rgb}{0 0 0.7}
\title{Quantum annealing applications, challenges and limitations for optimisation problems compared to classical solvers}

\author[1,*]{Finley Alexander Quinton}
\author[1, 2]{Per Arne Sevle Myhr}
\author[1]{Mostafa Barani}
\author[1,*]{Pedro Crespo del Granado}
\author[1]{Hongyu Zhang}

\affil[1]{Department of Industrial Economics and Technology Management, Norwegian University of Science and Technology, NO-7491 Trondheim, Norway.}
\affil[2]{Centre for Cosmology, Particle Physics and Phenomenology, Université Catholique de Louvain, B-1348 Louvain-la-Neuve, Belgium}

\affil[*]{quinton.f.alexander@ntnu.no; pedro@ntnu.no}

\begin{abstract}
Quantum computing is rapidly advancing, harnessing the power of qubits' superposition and entanglement for computational advantages over classical systems.
However, scalability poses a primary challenge for these machines.
By implementing a hybrid workflow between classical and quantum computing instances, D-Wave has succeeded in pushing this boundary to the realm of industrial use. 
Furthermore, they have recently opened up to mixed integer linear programming (MILP) problems, expanding their applicability to many relevant problems in the field of optimisation. 
However, the extent of their suitability for diverse problem categories and their computational advantages remains unclear. This study conducts a comprehensive examination by applying a selection of diverse case studies to benchmark the performance of D-Wave's hybrid solver against that of industry-leading solvers such as CPLEX, Gurobi, and IPOPT.
The findings indicate that D-Wave's hybrid solver is currently most advantageous for integer quadratic objective functions and shows potential for quadratic constraints. 
To illustrate this, we applied it to a real-world energy problem, specifically the MILP unit commitment problem. While D-Wave can solve such problems, its performance has not yet matched that of its classical counterparts.
\end{abstract}
\begin{document}

\flushbottom
\maketitle
%
%
\thispagestyle{empty}


\section{Introduction}

The field of Quantum Computing (QC) has seen great improvements over the last decade with companies like IBM and Google investing heavily in order to develop resources capable of solving challenging computational problems \cite{IBM_2021, google_2019}. Recently, Microsoft also announced their new topological qubit design \cite{microsoft2025interferometric}, showcasing a new path in obtaining more error-resistant, scalable quantum computers.  Although rapid progress coincides with modern times, the idea of using ``simple'' systems governed by Quantum Mechanics (QM) to perform computation was proposed already in the 1980s by both Richard Feynman \cite{Feynman_1982} and Paul Benioff \cite{Benioff_1980} independently. There are numerous ways to approach building a quantum computer, but the two main categories are gate-based and annealing-based QC. Just like bits in a classical computer, quantum computers use quantum bits, commonly referred to as qubits.
Over the years, many different technologies have been suggested and tested as physical implementations of qubits. The first attempts in the 90s used nuclear magnetic resonance \cite{first_qubit}, later, a plethora of different ideas has shown promise such as cold atoms \cite{cold_atoms}, trapped ions \cite{trapped_ions_qubits}, photon-based \cite{photon_qubits} and others \cite{quantum_dots,ladd2010quantum}. The current largest quantum annealers are pioneered by D-Wave \cite{adiabatic_quantum_dwave_book} and consist of loops of superconducting material \cite{devoret2004superconducting}. It has been shown that in specific cases, QC can be exponentially faster than Classical Computing (CC), such as for Shor's algorithm \cite{Shor_1997}, or the quantum Fourier transform algorithm \cite{QFT_parallel_circuits}.

The fundamental idea behind QC is to use quantum phenomena such as superposition and entanglement to solve computationally difficult problems. In addition to solving computationally expensive problems, QC may be suited for solving problems that classical computers cannot, such as simulating nature \cite{daley2022practical, grimsley2019adaptive}, as these systems are governed by the same rules of QM. It has been shown that for several NP-hard problems, where classical approaches fail to efficiently find a solution, QC can be a promising way to solve them \cite{np-hard_quantum, np_hard_quantum_opt, van2001powerful, Harwod2021}.

In gate-based QC, the classical, binary bits are replaced by controllable qubits capable of being in non-binary states. Adiabatic quantum annealers, however, function on a different principle. They prepare an ensemble of qubits in the ground state of some initial Hamiltonian, before allowing the system to evolve abiding by the adiabatic theorem \cite{adiabatic_theorem} to some final state encoding the desired solution. This principle will be further explained in \Cref{adiabatic_theorem}.

Although quantum annealers are easier to scale towards more qubits, they are far more limited in their applications due to the lack of controllability of the individual qubits. While in gate-based quantum computers, qubits can be controlled and manipulated with precise unitary operations, the annealing-based computers of D-Wave controls theirs with tunable biases and the coupling strengths between them. D-Wave's quantum annealers are not universal quantum computers, and therefore can not run arbitrary algorithms. They are, however, especially well suited for optimisation problems where scalability is one of the most determining factors in solving relevant, modern, real-world problems. For a more thorough overview of the synergies between operational research and quantum computing, see \cite{parekh2023synergies, abbas2023quantum}. 

The development of quantum computers has finally reached a point where they are comparable to classical computers for an ever-increasing diverse pool of problems, some of which we wish to present here. Several formulations of vehicle routing problems were proposed for utilising quantum computers in \cite{Harwod2021}. Syrichas \& Crispin \cite{syrichas2017} proposed a simplified but systematic approach to tune quantum annealing parameters for vehicle routing problems. Bernal et al. \cite{bernal2022perspectives} presented the perspective of quantum computing for chemical engineering and provided examples in computational chemistry and molecular simulation where QC may become relevant. Rosenberg et al. \cite{rosenberg2016building} developed a metaheuristic solver to solve large quadratic unconstrainted binary optimisation problems, which utilise D-Wave's quantum annealers. On the other hand, Glover, Lewis, and Kocheberger \cite{glover2018logical} analysed the logical and inequality implications of reducing problem size to use QC to solve larger problems. 

Quantum annealing may prove significantly beneficial for stochastic programming. In \cite{GU2010927} an algorithm based on the concepts of quantum theory was proposed and was applied to a stochastic job shop scheduling problem. In addition to real-world applications, several studies also investigate utilising quantum computing in classical optimisation algorithms, such as \cite{nannicini2022fast, LIU2010620}.



Although D-Wave's quantum annealers have been applied to solve problems such as rescheduling of urban railroads \cite{koniorczyk2023solving}, traffic flow optimisation \cite{neukart2017traffic}, and dynamic portfolio optimisation \cite{Mugel_2022}, the performance or capability of quantum annealers in solving different classes of optimisation problems compared with industrial leading classical solvers is not clear \cite{yarkoni2022quantum}. As pointed out in \cite{koniorczyk2023solving}, the current implementations of quantum annealing in optimisation are limited in size and not yet upscaled to real-world situations. Furthermore, the exploration of quantum annealing in scheduling problems is limited. Given the rapid evolution of QC, regular benchmarking becomes crucial to highlight emerging developments and showcase the current state of this highly promising field, as emphasised by \cite{abbas2023quantum}. Benchmarks like \cite{mohseni2024competitive} have shown that D-Wave can compete with classical solvers like Tabu Search \cite{tabu} and Simulated Annealing (SA) \cite{kirkpatrick1983optimization} for energy coalition formation and is ahead of gate-based applications like the Quantum Approximate Optimisation Algorithm \cite{farhi2014quantum}. However, no benchmark of industry-leading solvers has been conducted that evaluates their performance against that of D-Wave's hybrid solvers for different optimisation problem categories.

This research aims to address this gap by utilising the largest publicly available adiabatic quantum annealers, last updated in August 2023, to solve various optimisation problems, including integer/binary linear programming (BLP) with linear and quadratic constraints, integer/binary quadratic programming (BQP), and mixed-integer linear programming.
To assess performance, we systematically compare D-Wave's Advantage quantum annealers, as described in \cite{DWave_advantage}, with the state-of-the-art classical solvers CPLEX \cite{CPLEX}, Gurobi \cite{gurobi}, and IPOPT \cite{IPOPT}. Recognising the limitations of fully QC algorithms, we employ D-Wave's hybrid solvers \cite{dwave_hybrid_solver}. This approach seamlessly integrates the strengths of both classical and quantum computation, as explained in more detail in \Cref{hybrid_section}. This study serves as a valuable resource for researchers across various fields dealing with optimisation problems, enabling them to assess the suitability of current D-Wave quantum annealers for their respective applications.

The subsequent sections of the paper are organised as follows: \Cref{Theory} provides a brief introduction to the physical and technical background of quantum annealing, encompassing the adiabatic theorem and D-Wave's architecture in their treatment of constraints and the hybrid computing approach employed.  \Cref{Case_study} presents various case studies for different classes of optimisation problems. Then, \Cref{Unit commitment}  showcases the application of D-Wave's hybrid constrained quadratic model (CQM) solver \cite{dwave_hybrid_CQM} on an energy system problem formulated as Mixed-Integer Linear Programming (MILP).

\section{Quantum annealing}\label{Theory}
 Simulated Annealing (SA) is used for approximating the global optimal solution, often used when an approximate global solution is preferred over an exact local minimum \cite{kirkpatrick1983optimization}.  The fundamental principle of SA involves a two-phase process: initially ``heating'' the system, followed by a gradual ``cooling'' that guides it into its globally lowest energy state. It is important to emphasise that this ``heating'' and ``cooling'' are metaphorical concepts, as no physical heating or cooling is applied; rather, the entire process is mathematically simulated. The inspiration for SA comes from the physical annealing process often used to change the characteristics of different materials \cite{sheng2011catalyst}.
D-Wave utilises a quantum variation of this process, known as quantum annealing. In quantum annealing, the quantum mechanical phenomenon of tunnelling through a potential barrier, rather than the classical process of ``climbing'' over it, is utilised to escape local minima and potentially find a global one, as explained by Kadowaki and Nishimori (1998) \cite{kadowaki1998quantum}. This fundamental characteristic is pivotal in understanding the potential computational advantages of quantum annealers over their classical counterparts, particularly because the ``hill climbing'' process in SA can be time-consuming for a substantial subset of problems.

Quantum annealers exhibit narrower applicability compared to gate-based quantum computers and are, for example, unable to execute Shor's algorithm \cite{Shor_1997}.
By design, they are made to handle optimisation problems and are especially suited for quadratic unconstrained binary optimisation (QUBO) problems of the form
\begin{equation}
\begin{aligned}
\centering
\min_{x} \quad & \Bigl( \sum_{i} h_i x_i + \sum_{i<j} Q_{ij} x_i x_j\Bigr),\\
\end{aligned}
\label{QUBO}
\end{equation}
where $x\in \{0,1\}$ and $Q_{ij},h_i \in \mathbb{R}^{n}$.

QUBO problems are very closely related to the  \textit{Ising model} (Equation \Ref{class_ising})  and can be easily transformed into this form by applying an affine transformation $x \mapsto 2x - 1$, effectively mapping $x \in \{0,1\}$ to $\{-1,1\}$.  Under this transformation, the variables $x_i,x_j$ correspond to classical spin variables $s_i = \pm 1$ leading to the classical Ising spin system:
\begin{equation}
\centering
    E_{ising}(s) = \sum_i h_i s_i + \sum_{i>j}J_{i,j} s_i s_j,
\label{class_ising}
\end{equation}
where $J_{i,j}$ represents the spin interaction strength and $h_i$ external field.
Traditionally, the Ising model is used in the field of statistical mechanics where the variables represent the binary states of a particle that can either be ``spin down'' or ``spin up''. 

In quantum mechanics, these variables are replaced by Pauli operators $\sigma_{x,y,z}$. Specifically, spin states are often expressed by terms of the Pauli $\sigma_z$ matrix which correspond to spin projections along the z-axis. The Hamiltonian describing such a quantum system takes the form:
\begin{equation}
    H(\sigma) = {\frac{B(s)}{2}\Bigl(\sum_i h_i \sigma_z^{(i)} + \sum_{i>j}J_{i,j}\sigma_z^{(i)}\sigma_z^{(j)}\Bigr)},
\label{quantum_ising_hamiltonian}
\end{equation}
where $B(s)$ is a prefactor with dimensions of energy and  $J_{i,j}$ determines the interaction strength between neighboring spins. Since all the terms in this Hamiltonian involve $\sigma_z$ operators, they mutually commute. However, quantum annealers such as D-Wave introduce an additional transverse field involving $\sigma_x$ terms, which do not necessarily commute with $\sigma_z$. This non-commutativity is crucial for quantum annealing to leverage the adiabatic theorem \cite{adiabatic_theorem}, requiring an initial Hamiltonian with a dominant transverse field component as a starting point before gradually transitioning into the problem Hamiltonian.

\subsection{Adiabatic theorem}\label{adiabatic_theorem}

In quantum mechanics, operators are used to extract information about physical systems. The most relevant of these operators in our case is the Hamilton operator $H$, usually just referred to as the Hamiltonian. The Hamiltonian corresponds to the total energy of the quantum mechanical system. By assuming that the time-dependent part varies sufficiently slowly and that there is a big enough gap between the eigenstates of $H$, the solutions to the time-dependent Schödinger equation \cite{schrodinger1926} are obtained in terms of the eigenfunctions of the instantaneous Hamiltonian,
\begin{equation}
    H(t)\psi_a(t)=E_a(t)\psi_a(t),
\end{equation}
where $E_a$ is the energy of eigenstate $\psi_a$. The scale of ``sufficiently slow'' depends on the specific energy gap between the quantum states of the system and can range from less than a microsecond to a few seconds. If $H(t)$ varies slowly in time, a system initially in a non-degenerate state $\psi(t=t_0)$ with energy $E_a(t=t_0)$ will evolve into the corresponding state $\psi_a(t)$ with energy $E_a(t)$ at a later time $t$ without making any transitions between energy levels. This is known as the adiabatic theorem and is the fundamental principle utilised in quantum annealing. For a more detailed description of the adiabatic theorem and a complete proof thereof, see \cite{adiabatic_theorem}.

To perform calculations using a quantum annealer, the physical collection of qubits, hereafter referred to as the system, is prepared in an easy-to-solve initial Hamiltonian $H_I$. After the system has been prepared in $H_I$, external magnetic fields are applied to change the initial $H_I$ into a complex Hamiltonian $H_F$.
Suppose this process follows the adiabatic theorem described above. In that case, the system will remain in its ground state for the duration of the anneal, and the final state will be the solution of the complex Hamiltonian. As $[\sigma_x, \sigma_z]\neq 0$, $H_I$ and $H_F$ do not share a common set of eigenstates. As a result, the system must undergo quantum transitions to evolve from one eigenstate to another. If the two Hamiltonians did commute, the eigenstates would remain unchanged throughout the evolution, preventing any quantum dynamics from occurring. The procedure can slowly anneal the system from the quantum Hamiltonian (\Cref{quantum_ising}) to the solution of the classical Ising spin system shown in \Cref{class_ising}. After the annealing process, the collapsed spin states will present a low-energy solution. The annealing quantum Ising Hamiltonian is as follows:
\begin{align}
    H_\text{Ising} = &\underbrace{\frac{-A(s)}{2}\Bigl(\sum_i \sigma_x^{(i)} \Bigr)}_\text{Initial state Hamiltonian} + \underbrace{\frac{B(s)}{2}\Bigl(\sum_i h_i \sigma_z^{(i)} +\sum_{i>j}J_{i,j}\sigma_z^{(i)}\sigma_z^{(j)}\Bigr)}_\text{Final state Hamiltonian}.
\label{quantum_ising}
\end{align}

The annealing process begins at $t=0$ and ends at time $t = t_f$, where $A(0) \gg B(0)$ and $A(t_f) \ll B(t_f)$. At time $t=t_f$, the qubits have been de-phased to a classical system and the Pauli matrix can be replaced by the classical $s_i = \pm 1$, resulting in \Cref{class_ising}. 

\subsection{D-Wave architecture}\label{Dwave_architecture}

The quantum annealers developed by D-Wave are currently the most promising systems capable of utilizing quantum properties for computation. Since  2011, D-Wave has steadily improved the capabilities of their quantum processing units (QPU) by upgrading both the hardware through multiple generations of system topologies and their own developed software \textit{Ocean} \cite{dwave_ocean}, thereby allowing for remote problem solving on their quantum annealers. D-Wave quantum annealers use loops of superconducting material as qubits to build their QPUs. This approach fixes the physical layout, also known as the \textit(hardware graph), of the QPU.

 The most powerful QPU available is the \textit{Pegasus} topology \cite{pegasus_topo}. In the Pegasus topology, each qubit is coupled to 15 other qubits, which is a great improvement from the 6 couplings of the previous generation. It consists of a square lattice of unit cells which themselves contain 24 qubits. The Advantage QPU, used for the computations performed in this work, is a 16x16 lattice of unit cells, totalling $24\times16\times(16-1)=5760$ qubits. Due to the densely packed spatial arrangement of qubits and the physical manufacturing process, it is expected that some qubits may exhibit suboptimal performance. D-Wave estimates that approximately 5\% of the qubits in the Pegasus topology may have imperfections.  
 This dense graph structure of the Pegasus systems has allowed for a significant increase in the number of useful qubits, making the Advantage QPU the most promising system to solve the optimisation problems presented in \Cref{linear_section,quadratic_section,linear_quad_constr_section}. 
 
 Quantum devices like those from D-Wave are subject to several causes of errors, including thermal noise, control errors, flux bias drifts and decoherence, which can lead to wrong solutions. To mitigate some of these, D-Wave has implemented Drift Correction techniques \cite{perdomo2016determination} to compensate for slow fluctuations in qubit parameters by periodically measuring and adjusting flux biases to maintain computational accuracy. Additionally, D-Wave has demonstrated on the new Advantage-2 prototype Zero-Noise Extrapolation (ZNE) \cite{amin2023quantum} using controlled noise amplification and extrapolation to estimate error-free results. However, ZNE has not yet been implemented on commercially available systems. Generally, D-Wave uses statistical sampling by running multiple annealing cycles to generate a distribution of solutions, mitigating the noise introduced by these faulty qubits.
 
 Due to their physical implementation, current quantum annealers can only encode BLP and BQP problems, so terms and constraints have to be transformed into this form \cite{yarkoni2022quantum}. For detailed information on Pegasus and other D-Wave topologies, refer to\cite{chimera}. Additionally, the next generation of topology, called Zephyr \cite{zephyr}, has already been announced with even higher connectivity. When working with any of these topologies, the embedding of the problem onto the physical hardware is of high importance. While D-Wave does provide a large amount of functionalities to help in accomplishing this process, this is no easy task and requires significant knowledge of the architecture and the problem at hand.

\subsection{Constraints}\label{Dwave_constraints}
Once the Hamiltonian is set, the annealing process will result in a minimal energy configuration of the system. No restrictions regarding conditions or constraints on the binary variables can be enforced directly into this procedure. Constraints can, however, be accounted for by adding quadratic penalty terms directly to the Ising Hamiltonian, also known as Lagrangian relaxation \cite{fisher1981lagrangian}. By penalizing the unwanted solutions with large enough penalty factors, a cost is added to make some of the solutions unfavourable.

A linear equality constraint $P({\bf{x}})=\left(\sum_i^N a_i x_i -b\right)$ with binary variables $x_i=\{x_i: \forall i \in N\}$ can be written as
\begin{equation}
    \lambda \Bigl( \sum_i^N a_i x_i -b \Bigr)^2 \equiv \lambda P({\bf{x}})^2,
\label{lin_equal_constr}
\end{equation}
 where $b$ is some constant that has to be met. However, when confronted with a linear inequality constraint of the form $P({\bf{x}})= \left(\sum_i^N a_i x_i -b \right) \leq 0
$, auxiliary slack variables, ${\bf{y}}=\{y_j: \forall j \in J\}$ and their corresponding weights $w_j$ are used to transform inequality constraints into equality constraints \cite{hillier2001introduction} so that \Cref{lin_equal_constr} can be written as
\begin{equation}
    \lambda \Bigl( \sum_i^N a_ix_i + \sum_j^W w_j y_j -b \Bigr)^2 \equiv \lambda P({\bf{x}})^2.
\end{equation}
The number of slack variables $W$ needed can be determined based on the coefficients $a_i$ and $b_i$, with an upper limit of $W=b$. \newline 

The entire new QUBO, after adding all constraints, can then be written as 
\begin{equation}
    F({\bf{x}}) = \mathrm{Obj}({\bf{x}},Q) + \sum_{k} \lambda_k P_k({\bf{x}})^2,
\label{full_new_QUBO}
\end{equation}
with coupling values $Q$ and summing over all constraints $k$.
With this approach, constrained optimisation problems can be reformulated into a solvable QUBO. We note that
other approaches than Lagrangian relaxation have been considered and are actively researched to account for the constraints \cite{yu2021applying}. Furthermore, the penalty strength $\lambda$ in \Cref{full_new_QUBO} has to be chosen with care. Contrary to classical computing, there is a far more limited dynamic range of possible pre-factors and coupling strengths that can be applied. Too large values cause a tremendous amount of distortion of the distribution in the ground state. At the same time, too small values will not ensure the satisfaction of the constraint as breaking it could then still result in optimal, unfeasible solutions. The best penalty strength is thereby the lowest possible value that still satisfies the constraints. There is no consensus on how to consistently guarantee this behaviour, and only some guidelines can be found either empirically or algebraically \cite{yarkoni2021solving, neukart2017traffic}.

\subsection{State of the art of quantum annealing}
The most capable quantum annealers as of today are owned by D-Wave. Therefore, in this paper, we focus on analysing and comparing the performance of their most advanced system against that of classical computers. 
While D-Wave promotes the capabilities of their machines in solving optimisation problems with 1 million variables and 100,000 constraints \cite{DWave_advantage}, it is essential to note that these claims are strongly problem-dependent and not currently realizable simultaneously. The limit of 1 million variables is only valid for unconstrained problems with the possibility of adding constraints via penalties manually. For constrained models, the Advantage system can handle 500,000 variables and 100,000 constraints. These numbers, however, are only the case for D-Wave's hybrid solvers, which split the problem into different partitions, where some are solved classically and others on the QPU depending on how suited these sub-problems are for either of the systems. As an example, continuous variables are usually run on classical computers, as here quantum annealers are not suspected to provide any computational advantage \cite{ottaviani2018low}.
For pure quantum computation, without any partitioning, the capabilities of the Advantage system are significantly reduced and dependent on the exact embedding of the problem onto the QPU topology. Additionally, the larger the problem size, the ``smaller'' the logical qubits get, resulting in reduced connectivity and poorer solution quality. The maximum number of variables for a fully connected QUBO on the Advantage hardware is currently at 180, meaning the architecture can only support up to this amount (not considering any faulty qubits) of fully interacting logical variables \cite{Kuramata_2021}. For larger fully connected QUBOs, D-Wave uses hybrid workflows and partitioning of the problem into several sub-problems. 

\subsection{Hybrid computing}\label{hybrid_section}
One of the main issues with modern quantum computing and quantum annealers is their scalability. For most industrial problems, the current number of qubits is not sufficient to run purely via quantum methods. However, hybrid workflows pave the way to tackle this issue by identifying and decomposing the problem into several sub-problems. Hereby, the strengths of both classical computers and quantum annealers for suited tasks can be utilised similarly to GPUs used for training machine learning models. D-Wave's \textit{Leap Hybrid Solvers} \cite{dwave_hybrid_solver}
are cloud-based solvers and are the currently leading hybrid quantum annealing algorithm. Due to its proprietary nature, it acts like a black box, however, some generally known features are summarised in the following.
When identifying and decomposing the problem into several sub-problems, the \textit{LeapHybridCQMSolver}, designed for constrained quadratic models, automatically implements 
an appropriate penalty term for the constraint with suitable Lagrangian multipliers. D-Wave does not guarantee that this setting is optimal as these might differ for
certain problems.

After the preprocessing, a hybrid workflow \cite{dwave_hybrid_solver} is implemented utilising classical heuristics like simulated annealing and Tabu search \cite{tabu}. Simultaneously, the quantum hardware is used to run smaller binary quadratic unconstrained optimisation problems with the potential to boost the classical counterparts, noting that linear binary problems can easily be transformed into QUBOs. Furthermore, the hybrid solvers also take over the task of embedding the problem onto the quantum hardware and run the QUBO partitions several times, creating a sampleset of solutions that then can be incorporated into the hybrid workflow again. As both the quantum annealing and the integrated classical methods are heuristic, the entire output is also heuristic. Thereby the solver does not guarantee the optimal solutions but returns a sampleset of several results. The entire hybrid approach is probabilistic. Finally, the exact nature of these solvers is not known, however, D-Wave enables the creation of custom hybrid workflows which, if properly implemented, might outperform the \textit{LeapHybridSolvers}.
\section{Case study}\label{Case_study}

This section presents different case studies and their results. The exploration covers various classes of optimisation problems:
Integer/Binary Linear Programming (BLP), while also considering the influence of additional constraints of both linear and quadratic nature in \Cref{linear_section}, and Integer/Binary Quadratic Programming (BQP) in \Cref{quadratic_section}. 
To assess the performance of D-Wave's Advantage system for industrial use, we benchmark the solution quality and computational time of D-Wave's \textit{LeapHybridCQMSolver} \cite{dwave_hybrid_CQM} against results produced by the widely adopted classical solvers CPLEX \cite{CPLEX}, Gurobi \cite{gurobi}, and IPOPT \cite{IPOPT} (version 3.14.12). The classical solvers are executed on a dedicated computer cluster equipped with dual 3.5 GHz Intel Xeon Gold 5115 CPUs, each featuring 10 cores and 96 GB of RAM. Noting here that for the time comparison, we measured the total run time as outputted by the cloud service, which includes the time it takes to return the full sample-set of on average 100 sample solutions. From these solutions, we first filter out the feasible set before selecting the solution with the lowest energy value.
The time required for these two steps was not included in the computational time measurement for obtaining sufficient solutions. While these steps typically take less than 1s on average, their impact may only be considerable for very short total run times.

\subsection{Integer/Binary Linear Programming (BLP)}\label{linear_section}
\begin{figure}
     \centering
     \begin{subfigure}[b]{0.48\textwidth}
         \centering
         \includegraphics[width=\textwidth]{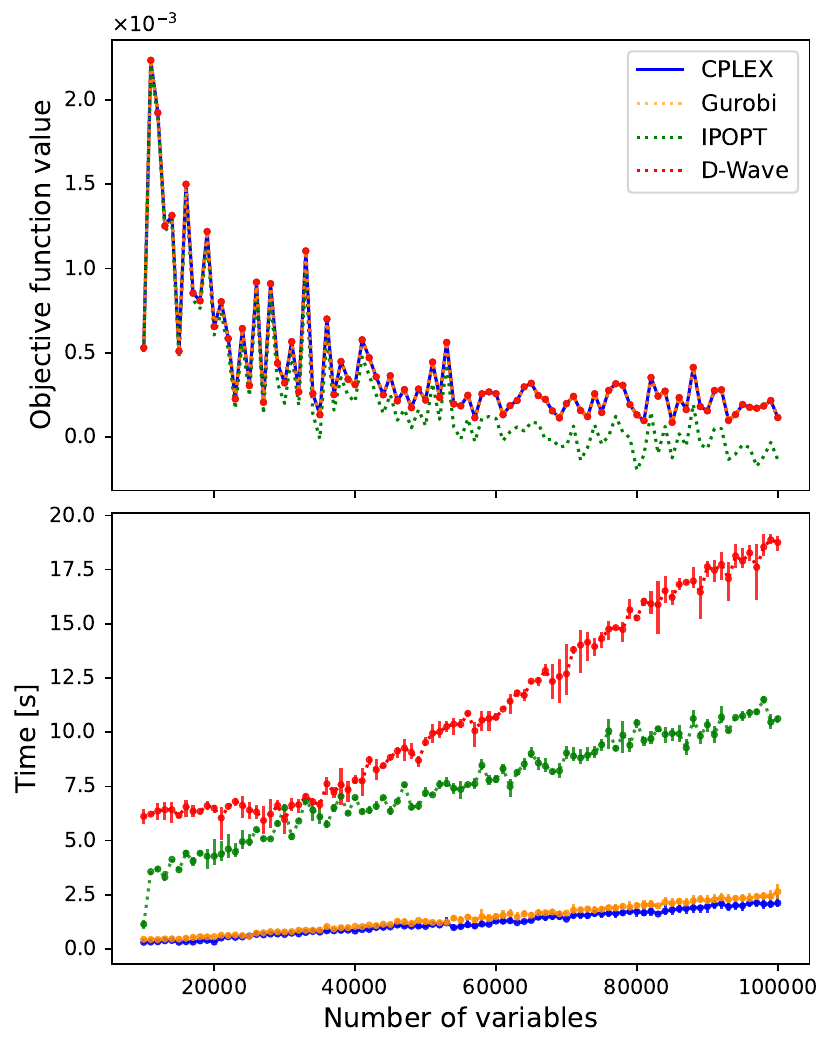}
         \caption{}
         \label{fig:linear_10_100k}
     \end{subfigure}
     \hfill
     \begin{subfigure}[b]{0.48\textwidth}
         \centering
         \includegraphics[width=\textwidth]{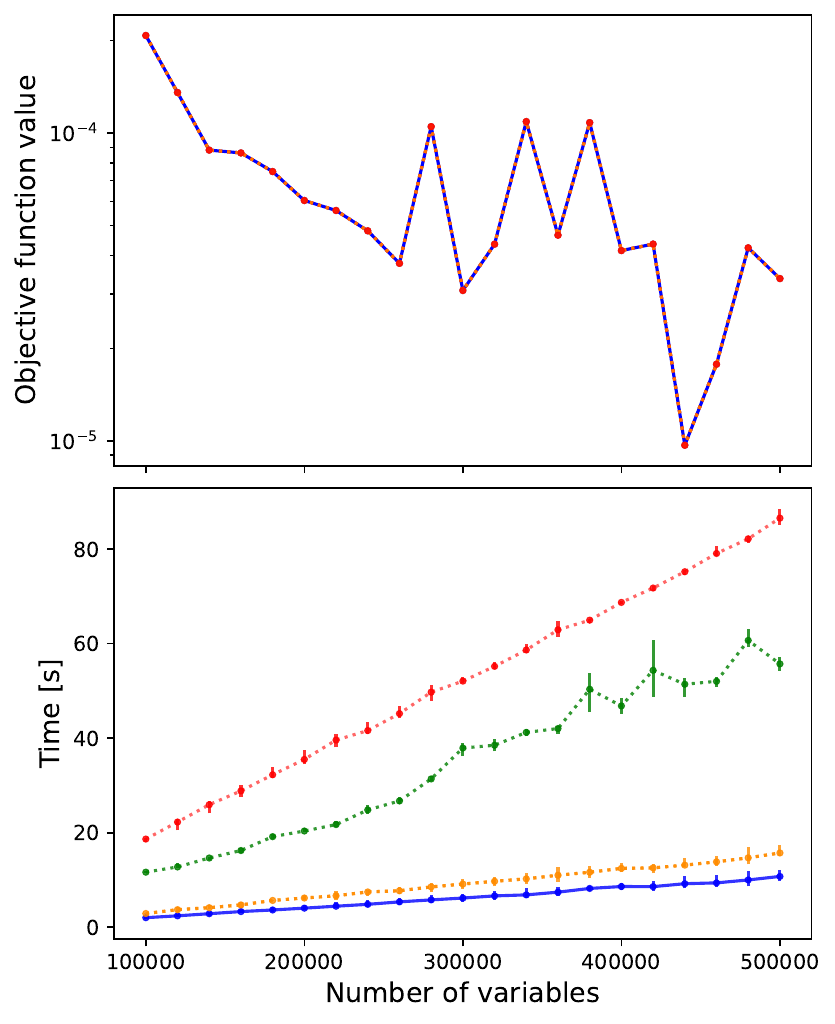}
         \caption{}
         \label{fig:linear_100_500k}
     \end{subfigure}
\caption{Mean objective function value and computational time comparison between CPLEX, Gurobi, IPOPT and D-Wave's hybrid solver for the \textbf{BLP} with a single linear constraint. The results are averaged over 5 runs and the minimum and maximum deviation from the mean value is shown by the vertical lines.}
\end{figure}

Initially, our focus is on exploring optimisation problems characterised by variables that are exclusively either integer or binary values, with linear objective functions and constraints. We consider how solution quality and execution times vary with the number of assigned variables for the hybrid CQM solver and compare its performance with the introduced classical ones. The problem is formulated as:
\begin{align}
\min \quad & \sum_{i=1}^{N} \mu_i x_i \label{linear_obj}\\
 \quad \textrm{s.t.} \quad & \sum_{i=1}^{N} x_i =  C,  \label{eq:c1}
\end{align}
where $\mu_i$ are random constants between (0,1) and $x_i \in \{0,1\}$  are binary variables. This restricts the sum over all binary variables to equal a fixed constant $C$. We first run the solvers with 10,000 to 100,000 binary variables. A comparison of the solutions and the computational time for this range of binary variables is shown in \Cref{fig:linear_10_100k}. Here, we average over 5 runs, each providing a sampleset of approximately 100 solutions from which the best feasible one is selected. As we experienced little deviation in the results we deemed higher statistics not to be necessary. Additionally, as access to quantum computers is still very limited, large statistics are not realisable. 
IPOPT is designed to handle large-scale nonlinear problems \cite{IPOPT} and is not suited for integer variables. As IPOPT approximates the binary variables with continuous ones with very small bounds around zero and one, the derived objective function value with these numbers can score a lower value than what is actually true. We excluded IPOPT's objective function values in \Cref{fig:linear_100_500k}, however, we kept this solver throughout this work as it shows important results in \Cref{quadratic_section}.
Looking at \Cref{fig:linear_10_100k}, D-Wave stays relatively constant in computation time of about 6s up to 38,000 variables, from whereon it increases more drastically when the number of variables is increased further.
Viewing the objective function value, we see that IPOPT behaves similarly to the other solvers for a low number of variables, but starts diverging from the rest as the number of variables increases. It can also be noted that the computational time of IPOPT increases rapidly compared to the other classical solvers. CPLEX and Gurobi start below 1s and gradually increase with an increasing number of variables. D-Wave finds the same optimal solution as Gurobi and CPLEX, resulting in three overlapping lines. The observed sawtooth pattern of the solutions is due to the randomness of the $\mu$ --values as they are newly drawn for each run with a specific seed. When comparing computational time in \Cref{fig:linear_100_500k}, the hybrid solver requires significantly more time as it approaches its scalability limit. This behaviour could be explained by the need for an extremely high amount of partitioning into subproblems as the number of variables increases \cite{yarkoni2022quantum}. Although the optimal solution has been found, D-Wave's hybrid solvers show no computational advantage.
\newpage
\subsection{Increasing the complexity by adding more constraints}
\begin{figure}
     \centering
     \begin{subfigure}[b]{0.44\textwidth}
         \centering
         \includegraphics[width=\textwidth]{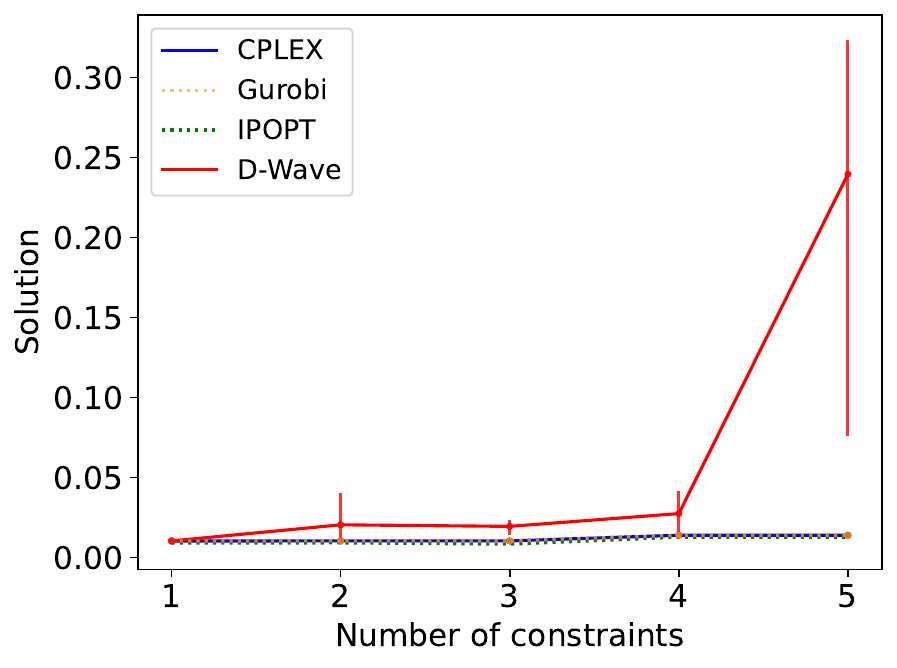}
         \caption{}
     \end{subfigure}
     \hfill
     \begin{subfigure}[b]{0.44\textwidth}
         \centering
         \includegraphics[width=\textwidth]{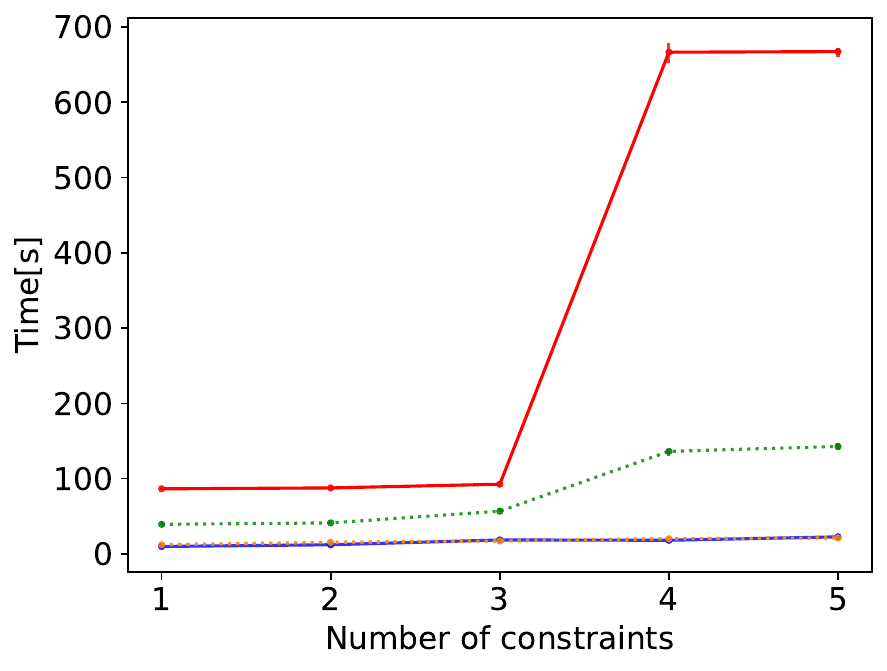}
         \caption{}
     \end{subfigure}
\caption{Mean objective function value \textbf{(a)} and computational time \textbf{(b)} comparison between CPLEX, Gurobi, IPOPT and D-Wave's hybrid solver for the \textbf{BLP} with increasingly more constraints. The results are averaged over 5 runs and the minimum and maximum deviation from the mean value is shown by the vertical lines.}
\label{fig:constraint_comp_500k}
\end{figure}

Next, we focus on determining whether D-Wave could continue to find the optimal solution at its scalability limit when increasing the complexity of the BLP (\Cref{linear_obj}) by adding additional constraints. As the constraints are added as penalty terms, we want to see if the solution quality can be maintained even though the computational time is significantly higher. We consider some arbitrary, although feasible, constraints listed below.
\begin{align}
&\sum_{i}^N x_i \geq C, \nonumber \\
&\sum_i^N x_i = \sum_j^N x_j, \quad i \in 2 \mathbb{N}, j \in 2 \mathbb{N} +1 , \nonumber\\
&\sum_i^{N/2} x_i \leq \sum_{i=N/2}^{N} x_i,\\
&\sum_i^N x_i = 0 , \quad i/5 \in \mathbb{N}, \nonumber\\
&\sum_i^N \mu_{i+1} x_i \leq \sum_i^N \mu_{i-1} x_i.\nonumber
\end{align}
The objective function value and computational time averaged over 5 runs, together with the minimum and maximum deviation as vertical lines, are shown in \Cref{fig:constraint_comp_500k}.
We observe that the solution quality stagnates with more constraints and in particular for the 5th added constraint a significant variability in the result with an approximate increase up to a factor of five. D-Wave's computational time is significantly higher reaching above 600s on the fully constrained problem. Furthermore, we ran the same problem with just 10,000 variables where D-Wave was able to find the optimal value with all 5 constraints added in every one of the 5 runs. This is a clear indication that D-Wave's CQM hybrid solver can not ensure optimal solutions at its variable limit. Although no thorough study on the best setting for the Lagrange multipliers was conducted, minor adjustments to their values showed no improvements, supporting our conclusion that D-Wave fails to provide optimal solutions at its variable limit with an increasing number of constraints.

\subsection{Integer/Binary Linear Programming with quadratic constraints }\label{linear_quad_constr_section}
\begin{figure}
     \centering
     \begin{subfigure}[b]{0.48\textwidth}
         \centering
         \includegraphics[width=\textwidth]{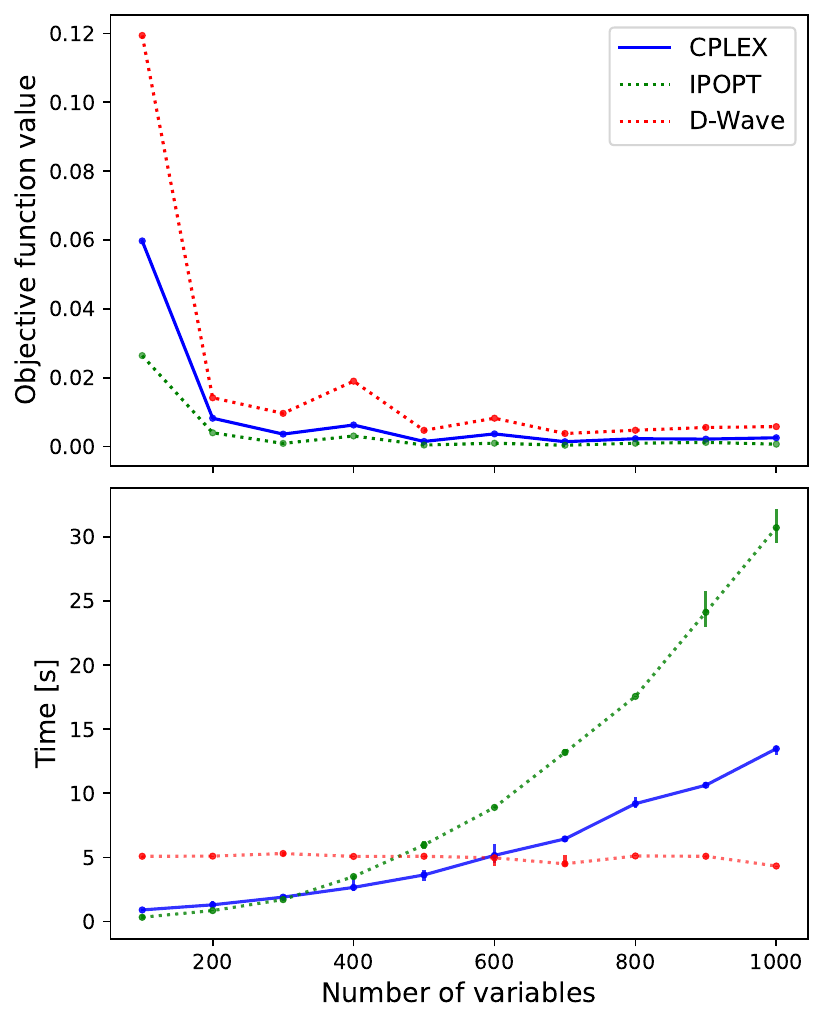}
         \caption{}
         \label{fig:quad_const_100-1000}
     \end{subfigure}
     \hfill
     \begin{subfigure}[b]{0.48\textwidth}
         \centering
         \includegraphics[width=\textwidth]{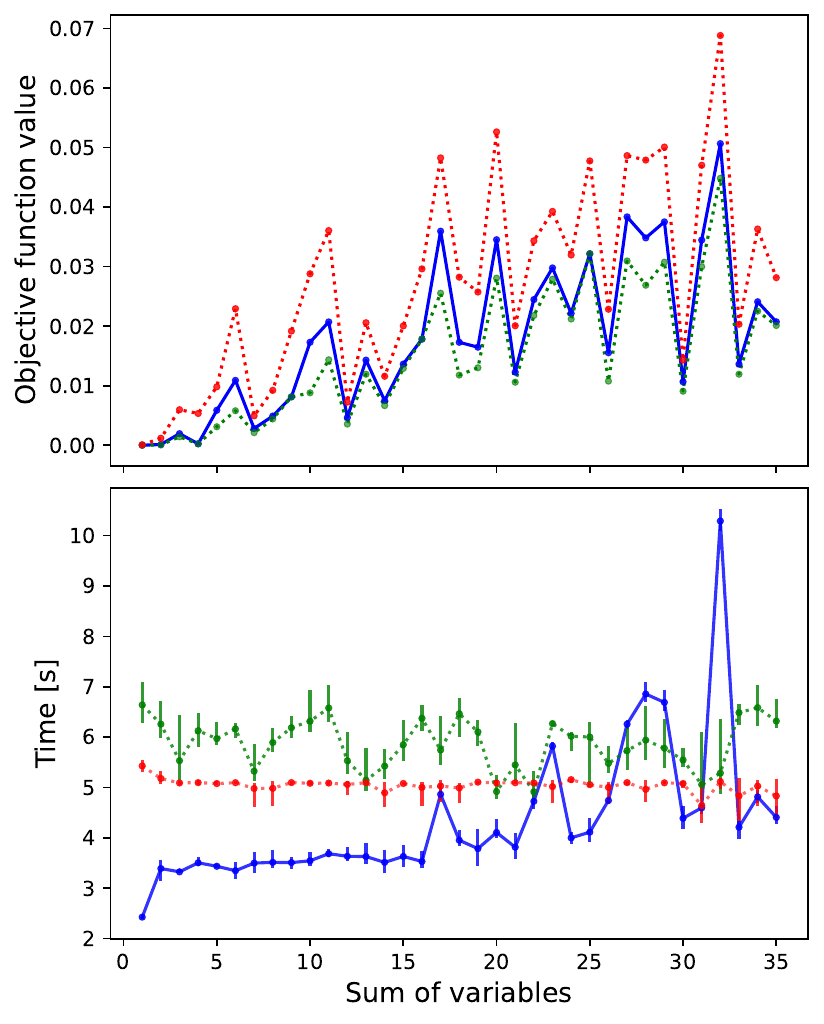}
         \caption{}
         \label{fig:quad_const_C_scale}
     \end{subfigure}
\caption{Mean objective function value (top) and computational time (bottom) comparison between two classical solvers and D-Wave's hybrid solver for the \textbf{BLP} with a single quadratic constraint. In \textbf{(a)} the complexity scales with the number of variables. In \textbf{(b)}, the variables are fixed to 500, whereas the complexity scales with increasing $C$ as stated in \Cref{eq:c1}. All results are averaged over 5 runs, and the vertical lines show the minimum and maximum deviation from the mean value. Gurobi's solve time is significantly larger than all other solvers and is thereby excluded here for visibility.}
        \label{fig:three graphs}
\end{figure}

In the previous section, all added constraints were linear. The examination, in this section, involves the consideration of a quadratic constraint while keeping the objective function linear.
To this end, we added another factor to the constraint from \Cref{eq:c1} so that it now reads
\begin{equation}
\sum_{i,j=1}^{N} x_ix_j \geq  C,  
\label{eq:cq1}
\end{equation}
where $x_{i,j} \in \{0,1\}$, resulting in $N^2$ terms. At first, we viewed the solution quality and computational time for an increasing number of variables, shown in \Cref{fig:quad_const_100-1000}, followed by a stepwise increase in the complexity of the constraint by increasing the value of $C$ in \Cref{fig:quad_const_C_scale}.

In the first scenario, we observe that D-Wave finds worse solutions than both CPLEX and IPOPT whereas IPOPT's value ranks lower due to the approximate representation of the binary variable,s as already mentioned in the previous section.

The solution obtained by CPLEX corresponds to the actual optimal solution, from which D-Wave is continuously approximately a factor of two off. Notably, although D-Wave does not find the optimal solution, the best solution found in every run is the same, resulting in no observable error bars. This indicates that D-Wave gets stuck in a local minima.
Nonetheless, D-Wave solves the problem faster than IPOPT from 500 variables and CPLEX from 600 variables onwards. Although D-Wave did not find the optimal solution a single time, the computational time is unaffected by the increase of variables.

We experience similar behaviour when stepwise increasing the value of the constraint. D-Wave is again slightly off the optimal solution, however, it stays relatively constant in computational time and solves the problem almost consistently faster than IPOPT. Although the solution is suboptimal, it can be seen in  \Cref{fig:quad_const_C_scale} that also for quadratic constraints with a linear objective function, the hybrid solver is not able to find the optimal solution in any run.
We note that we did not adjust the Lagrange multipliers and kept the ones chosen by default by the hybrid solver. We do not rule out that customised settings for the parameters could be beneficial to the quality of the solution.

\newpage
\subsection{Integer/Binary Quadratic Programming
(BQP)}\label{quadratic_section}

\begin{figure}
     \centering
     \begin{subfigure}[b]{0.44\textwidth}
         \centering
         \includegraphics[width=\textwidth]{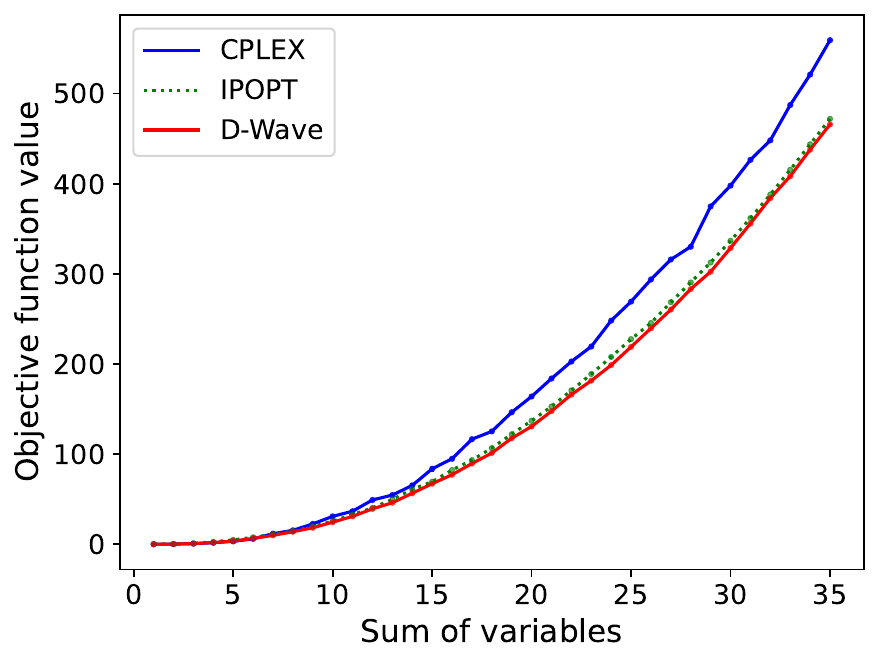}
         \caption{}
     \end{subfigure}
     \hfill
     \begin{subfigure}[b]{0.44\textwidth}
         \centering
         \includegraphics[width=\textwidth]{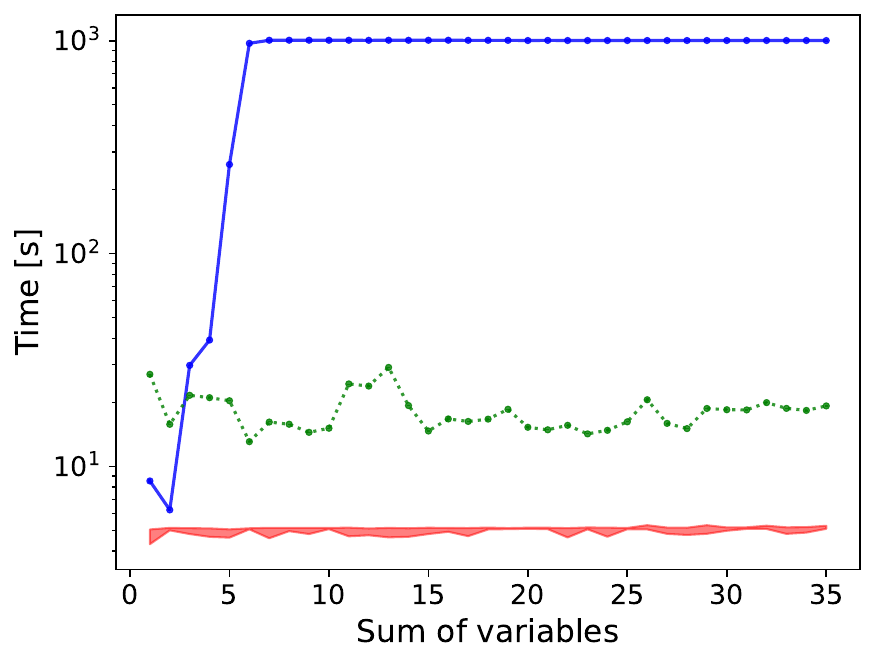}
         \caption{}
     \end{subfigure}
\caption{Comparison of the objective function value \textbf{(a)} and computational time \textbf{(b)} between D-Wave and the previously introduced classical solvers for \textbf{BQP} with 500 binary variables. The results are averaged over 5 runs. The run time is limited to 1000s, resulting in CPLEX not finding the optimal solution for increasing $C$.}
        \label{fig:quad_sol_comp}
\end{figure}

Binary quadratic problems are NP-hard \cite{binary_springer} and thus challenging to solve classically, even for a relatively small number of variables and constraints. As the structure of BQP resembles the \emph{Ising Hamiltonian}, \Cref{quantum_ising}, these problems are especially suited for quantum annealers  \cite{farhi2001quantum}. By multiplying with and summing over an additional set of binary variables $x_j$ in \Cref{linear_obj}, we acquire a BQP of the form 
\begin{align}
\min \quad & \sum_{i,j=1}^{N} \mu_{ij}  x_i x_j \label{quadratic_obj}\\
\quad \textrm{s.t.} \quad & \sum_{i=1}^{N} x_i =  C, \quad  x_i, x_j \in \{0,1\}.
\end{align}
To benchmark the different solvers for this BQP, we increase the complexity by incrementally increasing the value of the constant $C$.
Thereby, the complexity of the task is increased as the feasible solution space increases. This is preferred over a further increase of $N$ to avoid an unnecessarily large memory usage as the number of terms in the objective function increases with $N^2$.For the comparisons in this section, we choose a sufficiently large problem with $N=500$, although different sizes were tested, showing the same tendencies. The results are shown in \Cref{fig:quad_sol_comp}.

\begin{figure}[H]
    \centering
    \includegraphics[width=0.48\textwidth]{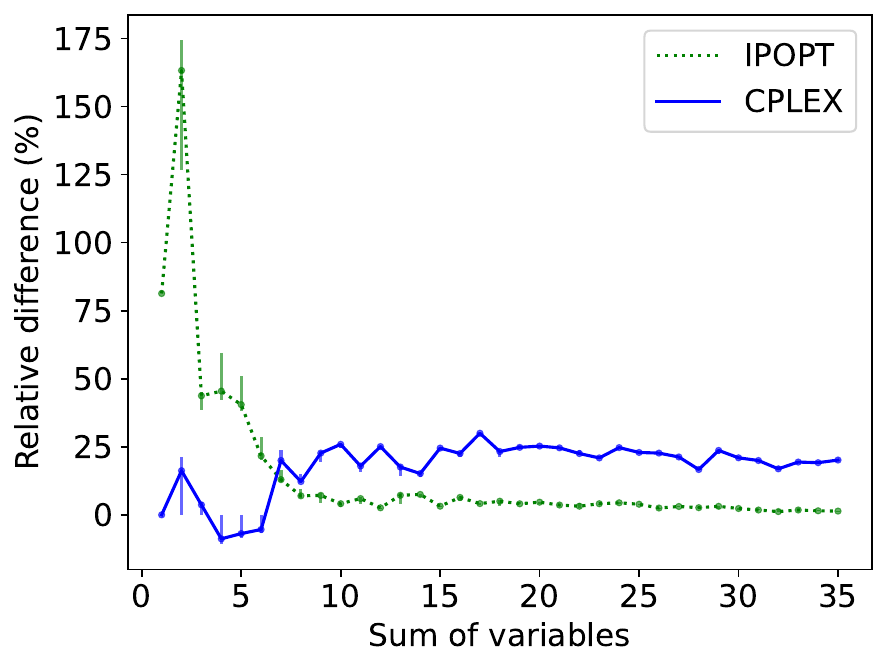}
    \caption[]{Mean relative difference of D-Wave's hybrid solution and those supplied by CPLEX and IPOPT for the \textbf{BQP}. The vertical lines indicate the difference between CPLEX's and IPOPT's mean solution to the maximum and minimum solution provided by D-Wave within all 5 runs.}
    \label{fig:logarithmic_diff}
\end{figure}

Considering both solution and run time, we observe a clear computational advantage of the hybrid solver to CPLEX. 
We cut CPLEX off after 1000s of runtime, resulting in sub-optimal solutions when the solution space increases. Gurobi performed even worse than CPLEX for this problem and was excluded from the figure for visibility reasons. While IPOPT shows comparable solutions and run time to D-Wave, the solutions are still not optimal. D-Wave finds consistently optimal solutions in less computational time, which again results in no error bars. To more clearly visualise this, we show the relative difference in solution quality between D-Wave's hybrid solver and the objective function value of both IPOPT and CPLEX in \Cref{fig:logarithmic_diff}.
We see from \Cref{fig:logarithmic_diff} that CPLEX performs well in the lower complexity region and that D-Wave does not find exclusively better solutions than CPLEX here; however, due to the capped computational time, D-Wave outperforms consistently for higher values of $C$. While IPOPT's relative difference becomes increasingly small, CPLEX, with a capped solve-time of 1000s, seems to settle at a deviation of $\sim 25\%$ from the optimal solution.
This shows that D-Wave's hybrid solvers are a suitable alternative for BQPs, even showing a computational advantage for this specific case. However, in the field of optimisation, there is a limited amount of problems that correspond to these pure BQPs. 

\newpage
\section{The unit commitment problem -- An energy system application}\label{Unit commitment} 
With the release of the LeapHybridCQM solver \cite{dwave_hybrid_CQM}, D-Wave opened itself up to a wider range of problems, such as Mixed Integer Linear Problems (MILP), with a wide spread of applications in various fields. To evaluate its performance, we benchmark D-Wave's current performance on a real-world Unit Commitment problem, a complex optimisation challenge in the power sector that plays a crucial role in power system operation and planning.
The Unit Commitment Problem involves determining the optimal scheduling of generating units---deciding when each unit should be turned on or off---over a given time horizon. The goal is to meet the forecasted load at minimum total production cost while satisfying a range of constraints, including power balance, reserve requirements, transmission limits, and individual unit constraints such as generation limits, minimum up/down times, and ramp rates \cite{knueven2020mixed}.
For this purpose, a well-established MILP formulation of the Unit Commitment problem is chosen \cite{Unit_github}, which consists of both continuous and binary variables while maintaining a fully linear structure in its objective function and constraints. The problem is tested at three different scales as defined in \Cref{Unit_table}. The objective function of a unit commitment model is as follows:
\begin{align}
        \text{min } \sum_{g \in G} \sum_{t \in T} \Bigg( \underbrace{c_g(t) + CP_g^1 \, u_g(t)}_{i} + \underbrace{\sum_{s = 1}^{S_g} \left( CS^s_g \delta^s_g(t) \right)}_{ii} \, \Bigg), \label{eq:obj} 
\end{align}
where the objective function consists of two components: (i) the production cost and (ii) the startup cost. $c_g(t)$ represents the cost of the power produced by thermal generator $g$ above its minimum output at time $t$, $CP_g^1$ represents the cost of operating at the minimum power output for generator $g$, while $u_g(t)$ is a binary variable that indicates the commitment status of generator $g$ at time $t$. 
In the final term, $CS_g^s$ represents the startup cost in category $s$ for generator $g$, where $s\in S_g$ represents the startup categories for thermal generator $g$, with $s=1$ corresponding to the hottest state (shortest downtime) and $s=S_g$ to the coldest state (longest downtime). The variable $\delta^s_g(t)$ indicates the startup in category $s$ for thermal generator $g$ at time $t$. These three terms are summed over all periods in time $t\in T$ and all generators $g\in G$. For a detailed explanation of this formulation, see \cite{Morales13,Unit_github}.

\begin{figure}[H]
     \centering
     \begin{subfigure}[b]{0.48\textwidth}
         \centering
         \includegraphics[width=0.98\textwidth]{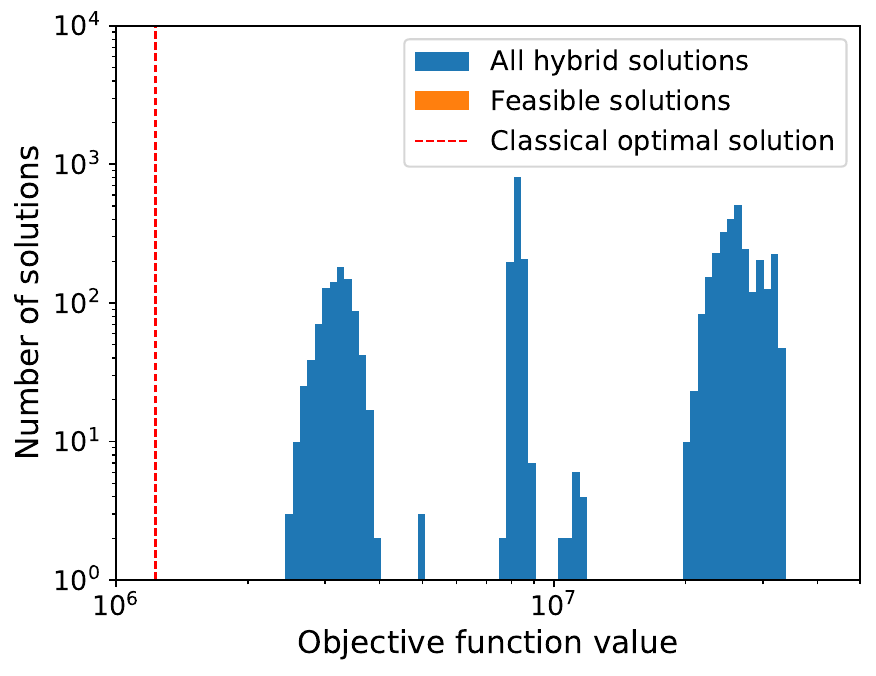}
         \caption{}
         \label{full_scale_default}
     \end{subfigure}
     \hfill
     \begin{subfigure}[b]{0.48\textwidth}
         \centering
         \includegraphics[width=\textwidth]{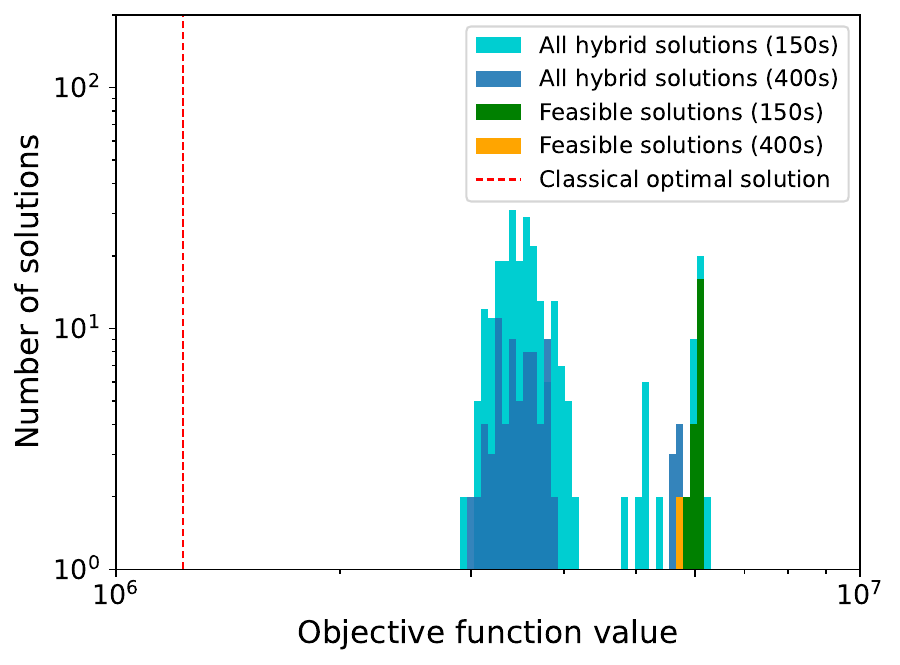}
         \caption{}
         \label{fig:full_scale_150_400s}
     \end{subfigure}
\caption{Solution distribution from D-Wave's LeapHybridCQM solver for the full-scale unit commitment problem. \textbf{(a)} shows the results of 50 runs with the solver's default chosen run time of 20s. \textbf{(b)} shows the solution for the same problem with 6 runs for an increased run time of 150s and 2 runs for one of 400s. Each run supplies a sampleset with on average 100 solutions each.}
        \label{fig:full_scale_total}
\end{figure}

\begin{figure}[H]
     \centering
     \begin{subfigure}[b]{0.49\textwidth}
         \centering
         \includegraphics[width=.985\textwidth]{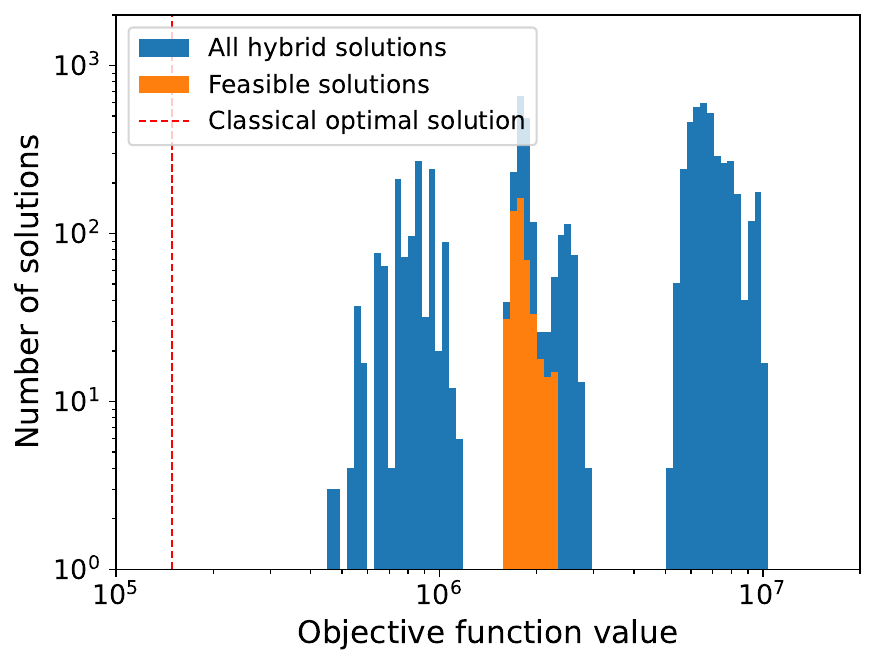}
         \caption{}
         \label{fig:unit_12h}
     \end{subfigure}
     \hfill
     \begin{subfigure}[b]{0.49\textwidth}
         \centering
         \includegraphics[width=\textwidth]{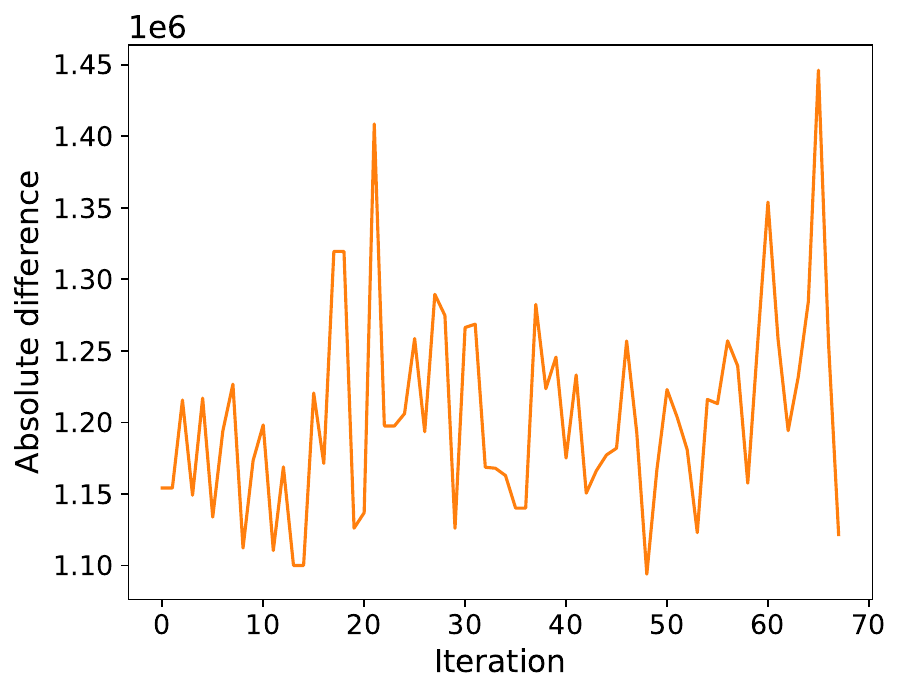}
         \caption{}
         \label{fig:12h_run_diff}
     \end{subfigure}
\caption{Solution distribution of the Reduced-1 unit commitment problem. \textbf{(a)} shows the objective function values of 70 runs (blue) given by D-Wave's LeapHybridCQM solver with an average of 100 solutions each. The feasible solutions are highlighted in orange. \textbf{(b)} shows the absolute difference between the optimal classical and the best feasible hybrid solution for each run for the Reduced-1 problem. Only one of the 70 runs did not obtain a feasible solution. }
        \label{fig:12_h_both}
\end{figure}

For the full-scale problem, the default pre-set minimum run time is set by D-Wave at 20s. In this limit, the hybrid solver was not able to find a single feasible solution, shown in \Cref{full_scale_default}. We therefore solved the problem 6 times with an increased minimum solve time of 150s and 2 times of 400s, shown in \Cref{fig:full_scale_150_400s}. For comparison, the near-optimal solution with a gap of 0.05\% was found by Gurobi in 180s with a value of $1.23 \times 10^6$. We can observe a clear increase in solution quality by the hybrid solver as in both cases, feasible solutions were found. The best feasible solution for the 150s case yielded, within 6 run,s an objective function value of $5.9 \times 10^6$, while two runs with 400s run time yielded a value of $5.7 \times 10^6$. This is a clear indication that the minimum run time chosen by the solver itself is not optimal and has to be adjusted specifically for the problem at hand, as a balance between computational time and solution quality.
Next, an investigation was conducted to observe how D-Wave would perform if the problem was downsized, as stated in \Cref{Unit_table}. First, the Reduced-1 problem with 12 time periods and 4 segments was examined, making it approximately one-fourth of the size of the initial problem. We kept the default minimum run-time setting of 10s, as this is already significantly longer than Gurobi's solve time of 1.34s for the optimal solution without a gap. The solution distribution of 70 runs is shown in \Cref{fig:unit_12h}, while the deviation of each run to the classical optimal solution is presented in \Cref{fig:12h_run_diff}. 

\begin{table}[H]
\centering

\begin{tabular}{lrrrr}
\toprule
 & Number of variables & Number of constraints & Time periods {[}h{]} & Segments \\ \hline
Full scale & 44544 & 42899 & 48 & 4 \\
Reduced-1 & 11136 & 10169 & 12 & 4 \\
Reduced-2 & 1418 & 1526 & 2 & 1\\
\bottomrule
\end{tabular}
\caption{\label{Unit_table}Parameters of the full scale and the two reduced versions of the Unit Commitment problem adjusted from \cite{Unit_github} and using the $rts\_gmlc$ data set therein.}
\end{table}

\begin{figure}
     \centering
     \begin{subfigure}[b]{0.49\textwidth}
         \centering
         \includegraphics[width=\textwidth]{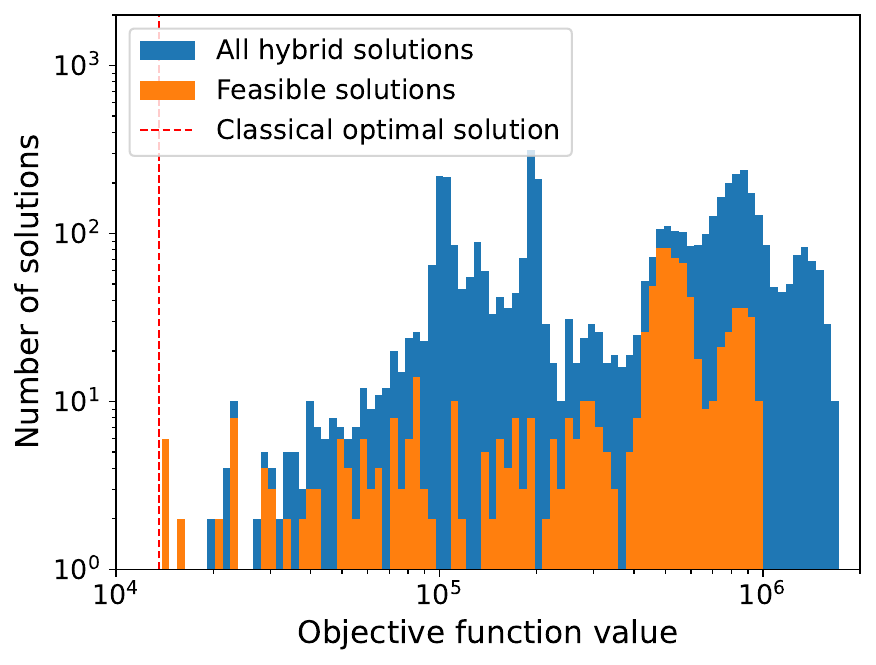}
         \caption{}
         \label{fig:unit_2h}
     \end{subfigure}
     \hfill
     \begin{subfigure}[b]{0.49\textwidth}
         \centering
         \includegraphics[width=.975\textwidth]{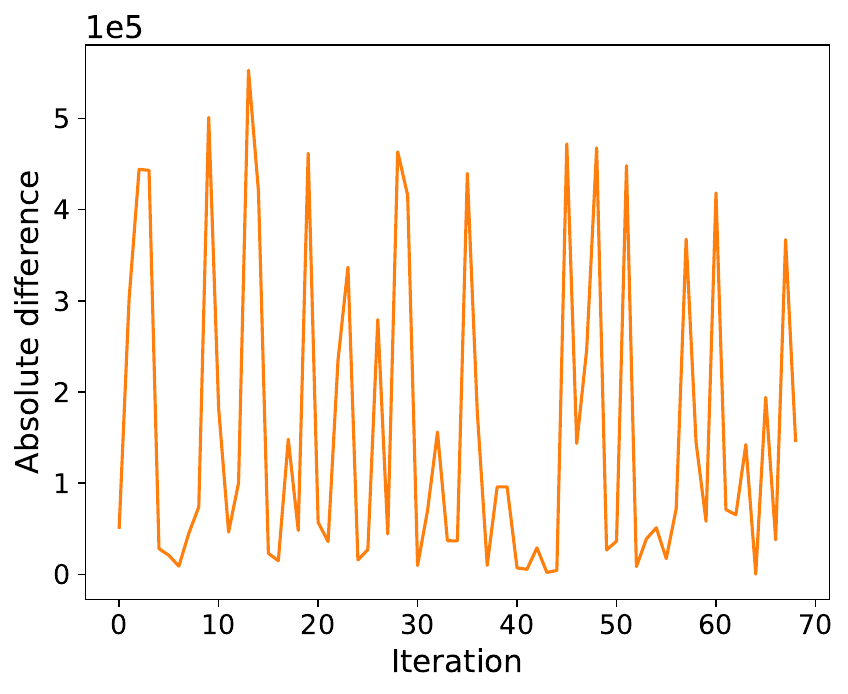}
         \caption{}
         \label{fig:2h_diff}
     \end{subfigure}
\caption{Solution distribution of the Reduced-2 unit commitment problem. \textbf{(a)} shows the objective function values of 70 runs (blue) given by D-Wave's LeapHybridCQM solver with an average of 100 solutions each. The feasible solutions are highlighted in orange. \textbf{(b)} shows the absolute difference between the optimal classical and the best feasible hybrid solution for each run for the Reduced-2 problem. A feasible solution was found in every run.}
\end{figure}

\newpage
Although D-Wave found at least one feasible solution in every run except for one, its best solution differs by approximately a factor of 10 from the optimal.
The Reduced-2 problem is reduced even further to just 2 time periods and a single segment of linearisation. The solution distribution over 70 runs is shown in \Cref{fig:unit_2h} and the deviation from the optimal classical solution in \Cref{fig:2h_diff}. 
The hybrid solver finds a feasible solution in every run with the best run being $\sim 2.4\%$ off from the optimal one. When comparing solve time, D-Wave takes on average 5s while Gurobi finds the best solution in less than 1s thereby showing no quantum computing advantage. We note that the problem in this reduced state is no longer representable of a real-world problem, as a 2h time frame is too short for a unit commitment problem. We conclude that D-Wave's hybrid solver is currently not in a state to effectively solve large-scale MILP unit commitment problems of this type, considering both computational time and solution quality. It is also worth mentioning that minor adjustments to some of the Lagrange multipliers for the constraints did not show any improvements.

\section{Conclusions}\label{Conclusion}
We provided an introduction to quantum annealing and conducted a benchmark of D-Wave's current generation of quantum-annealing-based hybrid solvers against state-of-the-art classical algorithms. Through a comparison of computational times and the corresponding best solutions across several case studies spanning a wide range of optimisation problems, we conclude that a state-of-the-art hybrid solver running on the most advanced quantum annealer has reached a point where it is competitive with the best classical approaches for a limited set of real-world problems.

For BLP, D-Wave's hybrid solver stays consistent with both CPLEX and Gurobi and finds near-optimal solutions for all our test cases. However, the computational time grows significantly when the number of variables increases. 

Although the hybrid-solver approach enables larger problem sizes than what is currently possible for full quantum approaches, such problems still have to be heavily partitioned into many subproblems and stuck back together, leading to increasing computational time \cite{yarkoni2022quantum}.

By increasing the complexity of the problem through systematically adding additional constraints, we conclude that the solution quality diverges more and more from the classical solvers for an increasing number of constraints at D-Wave's scalability limit. Additionally, the increase in run time is so drastic that the hybrid solver shows no benefit at this scale.
Although we did see some competitive computational times by D-Wave when handling quadratic constraints in \Cref{linear_quad_constr_section}, the hybrid solver was not able to find the optimal solution with a factor of approximately two off in most runs.
It is for the BQP that D-Wave's hybrid solver seems the most promising. Due to the mathematical structure of BQPs being so closely related to the Ising Hamiltonian of \Cref{quantum_ising}, D-Wave's hybrid solver performs better than all three classical solvers. The most competitive classical solver is IPOPT, but D-Wave does find slightly better solutions in a shorter amount of time. Compared with CPLEX, which due to its deterministic nature will find the optimal solution, D-Wave is superior in regards to computational time.
Regarding the MILP unit commitment problem, the benchmark underscored the importance of fine-tuning the minimum run time to enhance the solution quality. Despite various adjustments and reduced problem cases, D-Wave's \textit{LeapHybridCQMSolver} was unable to surpass Gurobi in terms of computational time or solution quality. As this specific problem has no quadratic binary instance in either the objective function or in any of the constraints, the case study indicates that these instances are currently necessary to potentially show a computational benefit. While it is promising that D-Wave can directly solve full MILP problems using their hybrid solvers, the computational advantage appears to be limited to BQP problems.

The findings presented in this paper indicate that D-Wave's hybrid solvers and Advantage quantum computers currently exhibit competitiveness with classical optimisation algorithms, but this holds true only for a limited range of problems. While D-Wave's hybrid solver performs well with binary quadratic problems, its competitiveness diminishes when dealing with non-binary, non-quadratic problems, or combinations thereof. Consequently, quantum annealers with broader capabilities have strong prospects for the future. The field of quantum annealing and quantum computing, in general, is undergoing rapid and promising advancements. 
Between 2020 to 2021 alone, D-Wave enhanced the performance of the Advantage machine, achieving an approximately eightfold increase in its ability to find solutions for Satisfiability problems and obtaining solutions for 3D lattice problems twice as fast \cite{d_wave_timeline}. They have already announced Zephyr for 2024 \cite{zephyr}. As classical solvers are experiencing regular performance updates, benchmarking research, as demonstrated in this paper, becomes imperative. 

\section*{Acknowledgements}
We acknowledge the support of the iDesignRES (Integrated Design of the Components of the Energy
System to Plan the Uptake of Renewable Energy Sources)1 project which has received funding from
the Horizon Europe research and innovation program under Grant Agreement No. 101095849.
\section*{Data availability}
The $rts\_gmlc$ dataset used for the unit commitment problem can be found in \cite{Unit_github}. All other datasets used and/or analyzed during the current study are available from the corresponding author upon reasonable request.




\section*{Author contributions statement}

F.A.Q, M.B., P.A.S.M., and P.C.G. conceived the work; F.A.Q and P.A.S.M. conducted the simulation; F.A.Q, P.A.S.M, and M.B. analysed the results; F.A.Q., P.A.S.M, M.B., and H.Z. wrote the original draft; All authors reviewed the manuscript.

\section*{Additional information}

The authors declare that they have no known competing financial interests or personal relationships that could influence the work reported in this paper.


\end{document}